\documentclass[twocolumn]{aastex61}

\usepackage{graphicx}	
\usepackage{amsmath}	
\usepackage{amssymb}	

\newcommand{\Msun}{\,{\rm M_\odot}}

\shorttitle{Most Lensed Quasars at $z>6$ are Missed by Current Surveys}
\shortauthors{Pacucci \& Loeb}

\begin{document}

\title{Most Lensed Quasars at $z>6$ are Missed by Current Surveys}

\correspondingauthor{Fabio Pacucci}
\email{fabio.pacucci@yale.edu}

\author[0000-0001-9879-7780]{Fabio Pacucci}
\affil{Yale University, Department of Physics, 
New Haven, CT 06511, USA}

\author[0000-0003-4330-287X]{Abraham Loeb}
\affiliation{Harvard-Smithsonian Center for Astrophysics,
Cambridge, MA 02138, USA}

\begin{abstract}

The discovery of the first strongly lensed $(\mu \approx 50)$ quasar at $z>6$ (J0439+1634) represents a breakthrough in our understanding of the early Universe. We derive the theoretical consequences of the new discovery. We predict that the observed population of $z > 6$ quasars should contain many sources with magnifications $\mu \lesssim 10$ and with image separations below the resolution threshold. Additionally, current selection criteria could have missed a substantial population of lensed $z > 6$ quasars, due to the contamination of the drop-out photometric bands by lens galaxies. We argue that this predicted population of lensed $z>6$ quasars would be misclassified and mixed up with low-$z$ galaxies. We quantify the fraction of undetected quasars as a function of the slope of the bright end of the quasar luminosity function, $\beta$. For $\beta \lesssim 3.6$, we predict that the undetected lensed quasars could reach half of the population, whereas for $\beta \gtrsim 4.5$ the vast majority of the $z > 6$ quasar population is lensed and still undetected. This would significantly affect the $z > 6$ quasar luminosity function and inferred black hole mass distributions, with profound implications for the ultraviolet, X-ray, and infrared cosmic backgrounds and the growth of early quasars.

\end{abstract}



\section{Introduction} \label{sec:intro}
The discovery of the first strongly lensed quasar at $z>6$ \citep{FAN_SCIENCE_PAPER} marks a breakthrough nearly two decades after the first theoretical prediction of their existence \citep{Wyithe_Loeb_2002}. \cite{FAN_SCIENCE_PAPER} report the detection of J0439+1634, a $z = 6.51$ quasar whose best-fit model predicts three images and a total magnification $\mu = 51.3 \pm 1.4$. This lensed quasar is the first example of this class detected during the reionization epoch \citep{BL01, Fan_2006}. 

\cite{Wyithe_Loeb_2002} suggested the possibility of a high fraction (up to $\sim 1/3$) of lensed sources among high-$z$ quasars, due to the large lensing optical depths reached at $z \gtrsim 6$. This fraction of the then-known population of high-$z$ quasars would have had their observed flux magnified by factors $\mu \gtrsim 10$. Thus far, none of the $\sim 150$ quasars observed at $z \gtrsim 6$ \citep{Banados_2018} have shown evidence for lensing.

As suggested by \cite{FAN_SCIENCE_PAPER}, this lack of lensed sources could be accounted for by a selection bias. The selection of $z \gtrsim 6$ quasars requires a non-detection (or at least a strong drop) in the drop-out band at shorter wavelengths than the Lyman break ($912 \, \mathrm{\AA}$). In their detection paper, \cite{FAN_SCIENCE_PAPER} point out that the presence of a lens galaxy along the line of sight of the lensed quasar injects flux into the drop-out bands. This effect is very relevant for partly resolved images, a likely occurrence for high-$z$ sources. This can thus constitute a relevant issue in the selection of $z \gtrsim 6$ lensed quasars. As the drop-out bands also contain transmission windows of the intergalactic medium (IGM), \cite{FAN_SCIENCE_PAPER} caution that great care is needed in separating the continuum of the lens galaxy from the transmission spikes of the IGM.

This new discovery has several key implications. First, it suggests the likely existence of a population of more modestly lensed quasars that are thus far undetected \citep{FAN_SCIENCE_PAPER}. Second, if this population of undetected quasars does exist, it could potentially impact in a very significant way the luminosity and mass functions of the earliest populations of supermassive black holes (BHs). In this regard, \cite{Wyithe_Loeb_2002} pointed out that the inclusion of a magnification bias in the high-$z$ quasar surveys could have a major effect on the abundance of massive halos ($M_h \gtrsim 10^{13} \Msun$), for which the mass function is very steep.

In this Letter we derive the theoretical consequences of the discovery of the first strongly lensed quasar in the epoch of reionization \citep{FAN_SCIENCE_PAPER}. Throughout the Letter we use the latest values of the cosmological parameters from the \cite{Planck_2018}.

\section{Theoretical Framework} 
\label{sec:theory}

We employ the formalism developed in \cite{Pei_1995} to analytically compute the magnification probability distribution, $P(\mu)$, due to cosmologically distributed galaxies.
We define $\mu$ as the total magnification of a source at a redshift $z_s$, due to lenses at redshift $z'$. The probability distribution of magnifications larger than $\mu$ is
\begin{equation}
P(>\mu) = \int_{\mu}^{\infty} P(\mu') \mathrm{d\mu'} \, .
\end{equation}

Following \cite{Pei_1993, Pei_1995}, we distinguish between the magnification $\mu$, relative to a filled beam in a smooth Universe, and the magnification $A$, relative to an empty beam in a clumpy Universe. Calling $\bar{A}$ the mean magnification, the two variables are related by $\mu \equiv A/\bar{A}$.
We define the moment function $Z(s|z_s)$ as

\begin{equation}
\begin{split}
Z(s|z_s) = & \int_{0}^{z_s} \mathrm{d}z' \int_{1}^{\infty} \mathrm{d}A \rho(A, z'|z_s) \times \\
& \times (A^s -1 -0.4\ln{(10)}A +s) \, .
\label{eq:moment}
\end{split}
\end{equation}
The quantity $\rho(A, z'|z_s)$ is the mean number of lenses in the range ($z'$, $z' + dz'$) with magnifications in the range ($A$, $A + dA$) for a given source at a redshift $z_s$.
The probability distribution function for $\mu$ reads
\begin{equation}
P(\mu) = \mu^{-1} \int_{-\infty}^{+\infty} \mathrm{d}s \exp{[-2\pi i s \ln{\mu} +Z(2 \pi i s|z_s)]} \, .
\label{eq:pdf}
\end{equation}
With some analytical adjustments, this integral can be efficiently computed via a fast Fourier transform algorithm. 

Assuming that the population of lenses is composed of galaxies with flat rotation curves modeled as a truncated singular isothermal sphere, their surface mass density profile can be written as
\begin{equation}
\Sigma(R) = \frac{\sigma_v^2}{2GR} \left( 1 + \frac{R}{R_T} \right)^{-2} \, ,
\end{equation}
where $\sigma_v$ is the line-of-sight velocity dispersion, $R$ is the projected radius, $R_T$ is the radius containing half of the projected mass, and $G$ is the gravitational constant. The total mass is finite and equal to $M_T = \pi \sigma_v^2 R_T/G$. 
For such a population of lenses, calling $\delta(x)$ the Dirac delta function
\begin{equation}
\rho(A, z'|z_s) = 2 \tau(z'|z_s) \int_{0}^{\infty} \mathrm{d}f f \delta[A-A(f)] \, .
\label{eq:rho}
\end{equation}
The coefficient $\tau(z'|z_s)$ is the strong lensing optical depth between source and lens galaxy at $z'$ (see \citealt{Pei_1995}). 
The factor $f$ is defined as $f = l/(a_{\mathrm{cr}}R_T)^{0.5}$, where $l$ is the unperturbed impact parameter, and $a_{\mathrm{cr}}$ is the critical impact parameter for double images.
In Eq. \ref{eq:rho}, $A(f) = A_+(f)$ for $f>1/r$ and $A(f) = A_+(f) + A_-(f)$ for $f \leq 1/r$, with
\begin{equation}
A_{\pm}(f) = \frac{1}{2} \left( \frac{f^2 \pm fr +2}{f[(f \pm r)^2 +4 ]^{1/2} } \pm 1 \right) \, ,
\end{equation} 
\begin{equation}
r = \left[ \frac{R_T c H_0}{4 \pi \sigma_v^2 D(z_s, z')} \right]^{1/2} \, .
\end{equation}
Here, $D(z_s, z')$ is the angular diameter distance between source and lens at $z'$, and $c$ is the speed of light.
The size parameter $r$ has dimensions of $\mathrm{length}^{-1/2}$ and is usually expressed as a function of the dimensionless parameter $F = 3 \Omega_G/[2 r^2 D(z_s, z')]$, with $\Omega_G$ the density parameter of galaxies. Following \cite{Pei_1995}, we assume $F \sim 0.05$ and we check that our results are unchanged throughout the range $ 0.01<F < 0.1$, which covers the full domain of interest. 

\subsection{Distribution of lens galaxies}
The theoretical framework described thus far to make predictions for the lensing probability $P(>\mu$) has to be supplemented with a realistic cosmological distribution of foreground ($z \lesssim 6$) galaxies, assumed to be early type (E/S0) and distributed uniformly in space.
The ultraviolet (UV) luminosity function for galaxies is modeled as a simple Schechter function \citep{Schechter_1976}:
\begin{equation}
\Phi(L) = \frac{\Phi_{\star}}{L_{\star}} \left( \frac{L}{L_{\star}} \right)^{\alpha_g} \exp{\left(-\frac{L}{L_{\star}} \right)} \, ,
\end{equation}
where $L_{\star}$ is the break luminosity, $\Phi_{\star}$ is the number density of galaxies of luminosity $L_{\star}$, and $\alpha_g$ is the slope of the faint end. Several surveys (e.g. \citealt{Schmidt_2014, Coe_2015}) suggest that the Schechter function correctly models the distribution of galaxies up to $z \sim 6$, and possibly up to $z \sim 8$ \citep{Bouwens_2014}.
We employ previous results on the UV luminosity function for galaxies, which are customarily divided into two large redshift ranges. For $z < 1$ we follow \cite{Beifiori_2014}, while for $z \gtrsim 1$ we follow \cite{Bernardi_2010} and \cite{Mason_2015}. As shown by previous studies (e.g., \citealt{Wyithe_2011, Mason_2015}) most of the lensing optical depth for sources at $z \gtrsim 6$ originates from lens galaxies at $z \lesssim 1.5$.
The dependence of the velocity dispersion $\sigma_v$ on the magnitude of the galaxy is modeled assuming the Faber-Jackson relation \citep{Faber_Jackson_1976}, i.e. $L \propto \sigma_v^4$.
As the peak of star formation activity is reached at $z \sim 2$ \citep{Madau_2014}, the stellar velocity dispersion $\sigma_v$ is expected to increase with redshift, at least for $z \lesssim 3$. This redshift range includes the population of lens galaxies that contribute the most to the lensing optical depth. We model the redshift evolution of the stellar velocity dispersion as $\sigma_v (z) \propto (1+z)^{\gamma}$. Following \cite{Beifiori_2014}, we use the value $\gamma = 0.18 \pm 0.06$, indicating a mild evolution. This value is consistent with other studies (e.g., \citealt{Mason_2015}), but smaller than what found by \cite{vandesande_2013}. As discussed in \cite{Mason_2015}, measurements of velocity dispersions at $z \gtrsim 0.5$ are very difficult, and large uncertainties in lens models result from a lack of knowledge of how the dark matter contained in galaxies (traced by $\sigma_v$) evolves with redshift.

\section{RESULTS} 
\label{sec:results}
We are now in a position to make theoretical predictions for the population of lensed $z > 6$ quasars.
Our model for the population of sources assumes a double power-law shape for the $z > 6$, quasar luminosity function, with a faint-end slope $\alpha = 1.3$ \citep{Manti_2017} and a variable bright-end slope, as discussed in the following.

\subsection{The population of lensed $z>6$ quasars} 
\begin{figure}
\includegraphics[angle=0,width=0.5\textwidth]{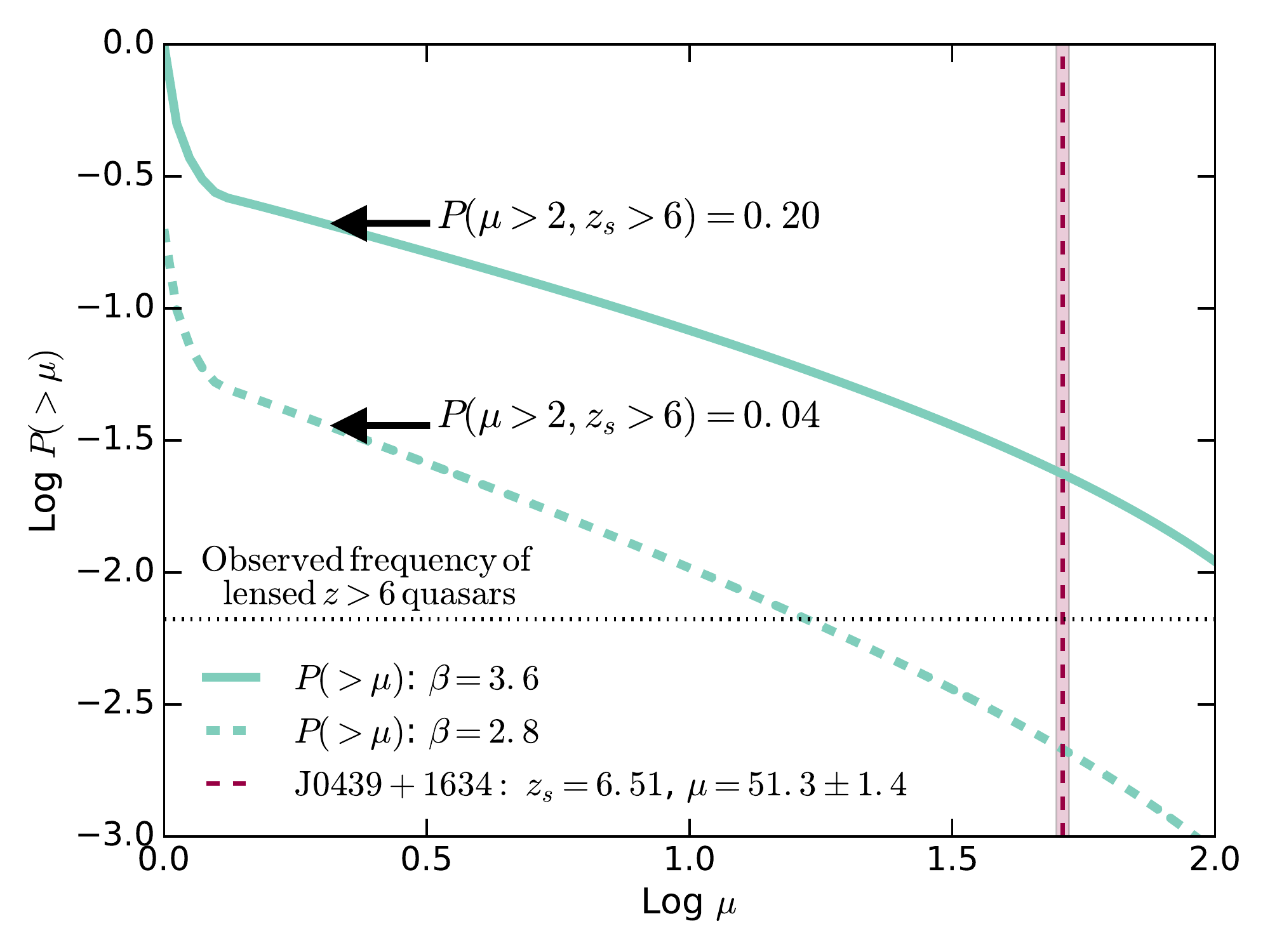}
\vspace{-0.4cm}
\caption{The probability distribution function $P(>\mu, z_s = 6.51)$ is shown for $\beta = 2.8$ and $\beta = 3.6$, and for magnification values between $\mu = 1$ and $100$. The dashed vertical line indicates the magnification factor for the lensed quasar reported by \cite{FAN_SCIENCE_PAPER}, with the $1\sigma$ uncertainty level shown as a shaded region. Note that $P(\mu > 50) = P_{\rm obs}$ for $\beta \approx 3.2$.}
\label{fig:theory}
\end{figure}

The resulting probability distribution function $P(>\mu)$ for a source at $z = 6.51$ is shown in Fig. \ref{fig:theory}.
Defining $P(\mu > 2, z_s > 6) \equiv P_0$ as the probability of strong lensing ($\mu>2$, leading to multiple images, e.g. \citealt{, Comerford_2002}), the plot shows the result for $P_0 = 0.20$, valid if the slope of the bright end of the quasar luminosity function $\beta$, $\Phi(L) \propto L^{-\beta}$, is $\beta = 3.6$ \citep{Yang_2016}, and for $P_0 = 0.04$, valid if $\beta = 2.8$ \citep{Jiang_2016}. 
Note that for $\beta \gtrsim 4.5$ we obtain $P_0 \gtrsim 0.92$. Very recently \cite{Kulkarni_2018} predicted the bright-end slope of the $z \sim 6$ quasar luminosity function to be as high as $\beta \approx 5.05^{+1.18}_{-0.76}$.

In our case, the observed frequency of lensed $z > 6$ quasars is $P_{\rm obs} \sim 1/150 \approx 7\times 10^{-3}$ \citep{Banados_2018}, with one source at $z_s = 6.51$ having a magnification factor $\mu \approx 51$. If we assume that this source is drawn randomly from a population smoothly distributed in $\mu$, we require that $P(\mu > 50) \geq P_{\rm obs}$. Depending on the value of the slope $\beta$, the predictions for $P(\mu > 50)$ vary, and we obtain $P(\mu > 50) = P_{\rm obs}$ for $\beta \approx 3.2$ (see Fig. \ref{fig:theory}).

A discrepancy between the frequentist approximation of the probability and the theoretically computed value can be explained by a magnification bias \citep{Turner_1980}. The corresponding factor ${\cal B}$ is employed in surveys to account for the fact that lensed sources are brighter than the unlensed population from which they are drawn. When a source with an observed magnitude $m$ is lensed, the probability of detecting it is $\sim {\cal B}(m)$ times higher than the probability of detection without the lensing effect. Following \cite{Schneider_2006}, we express the magnification bias ${\cal B}(m)$ as
\begin{equation}
{\cal B}(m) = {\cal N}(m)^{-1} \int \mathrm{d\mu} \frac{dP}{d\mu} {\cal N}(m+2.5\log\mu ) \, ,
\end{equation}
where ${\cal N}(m)$ is the number count of sources with magnitude $m$. Assuming a typical power-law form for the number counts in flux ${\cal N}(F) = d{\cal N}/dF \propto F^{-\alpha}$, the corresponding expression in magnitude is ${\cal N}(m) \propto 10^{0.4(\alpha - 1)m}$.
Assuming $\alpha = 1.3$ as the faint-end slope of the $z \sim 6$ quasar luminosity function, the magnification bias ${\cal B}(m)$ varies between a value of $\sim 25$ at $m_{\mathrm{AB}}=19$ and a value of $\sim 2.5$ at $m_{\mathrm{AB}}=23$.
For J0439+1634 ($m_{\mathrm{AB}} = 21.04 \pm 0.01$) we thus expect a bias factor of $\sim 10$.
\begin{figure}
\includegraphics[angle=0,width=0.48\textwidth]{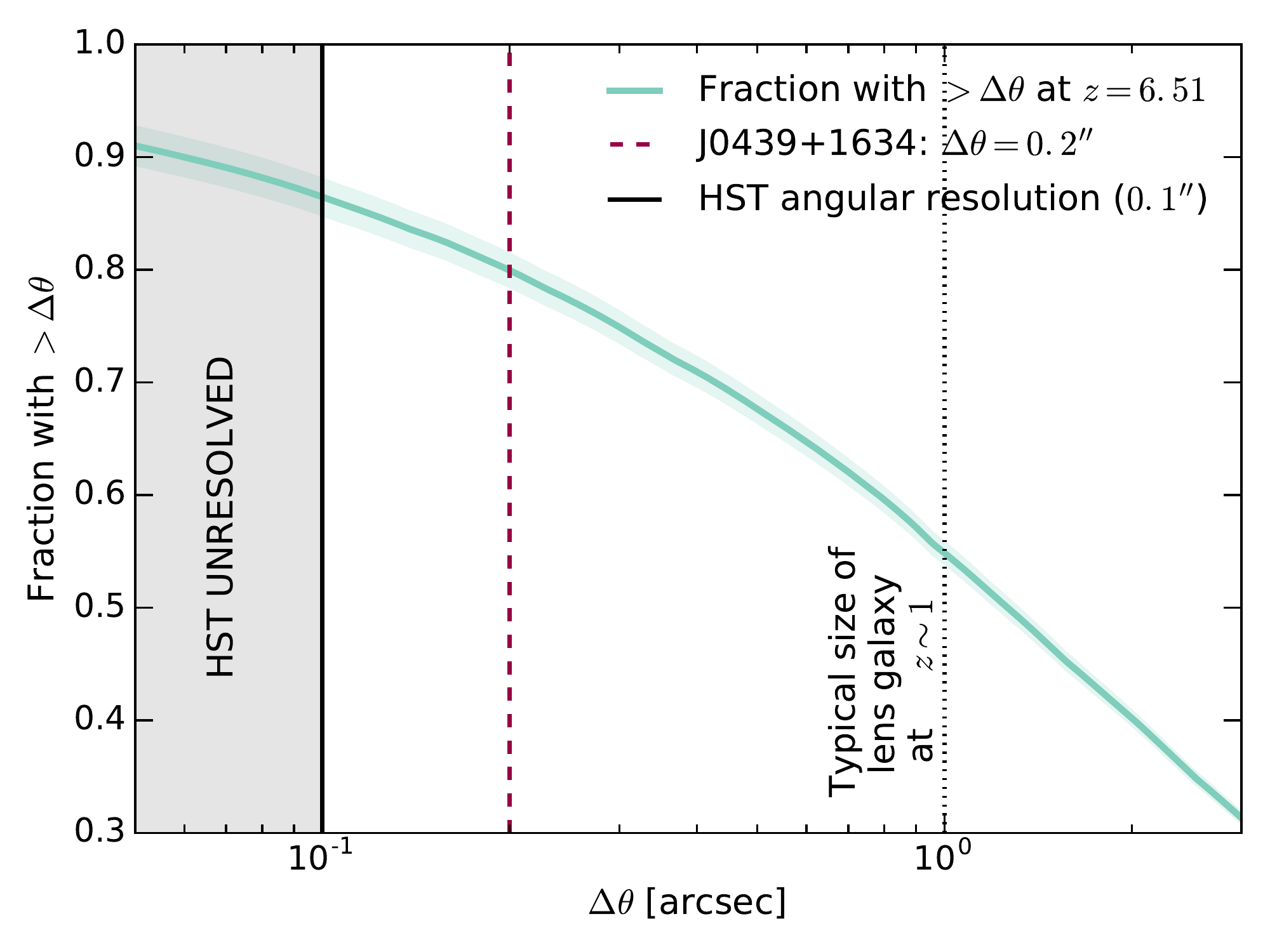}
\vspace{-0.4cm}
\caption{Distribution of image separations at $z = 6.51$, obtained assuming a Press-Schechter distribution of galaxies, following \cite{BL_2000}. The shaded area indicates the $1\sigma$ uncertainty due to the distribution of velocity dispersion in galaxies $\sigma_v$. The angular separation for J0439+1634 ($\sim 0.''2$), the HST angular resolution ($\sim 0.''1$) and the typical angular size of lens galaxies at $z\sim 1$ ($\sim 1''$) are also shown for reference.}
\label{fig:bias}
\end{figure}
While the detection of J0439+1634 was serendipitous and not necessarily representative of the population from which it was drawn, a comprehensive probabilistic analysis with more examples can inform us about the value of the slope $\beta$, once the magnification bias factor is taken into account.
For any value of $\beta <3.6$, our predictions for the strong lensing probability $P(\mu>2)$ are lower than the value estimated by \cite{Wyithe_Loeb_2002}, i.e. $P(\mu>2) \sim 0.3$. In this regard, \cite{Mason_2015} pointed out that early works (e.g., \citealt{Wyithe_Loeb_2002, Wyithe_2011}) might have overestimated the strong lensing optical depth. The values suggested by \cite{Mason_2015}, $P(\mu>2) \sim 3\%-15\%$, are consistent with our estimates.

\subsection{An undetected population of $z > 6$ quasars}
Figure \ref{fig:theory} implies that, \textit{independently on the value $\beta$ of the bright end of the quasar luminosity function}, the probability of observing quasars with a magnification $ \mu \lesssim 10$ is higher than the observed frequency of lensed $z > 6$ quasars. This indicates the theoretical possibility that some of the observed $z > 6$ quasars are magnified  with factors $\mu \lesssim 10$.
High-resolution imaging spectroscopy could potentially reveal whether or not a quasar is lensed. Furthermore, diagnostics such as the quasar proximity zone (e.g., \citealt{Eilers_2017}) allow an estimate of the intrinsic luminosity of the quasar. However, it is crucial to note that the maximum image separation for J0439+1634 is $\sim 0.''2$ \citep{FAN_SCIENCE_PAPER}, close to the highest-resolution limit obtainable with the \textit{Hubble Space Telescope} (HST; see Fig. \ref{fig:bias}). Therefore, it is a clear possibility that many of the detected $z>6$ quasars are actually lensed, with image separations below the resolution threshold. Following \cite{BL_2000}, in Fig. \ref{fig:bias} we show the distribution in angular separations of lensing images for $z=6.51$, $\Delta\theta = 8\pi (\sigma_v/c)^2 D(z_s,z')/D(z_s,0)$, where $D(z_s,0)$ is the angular diameter distance between source and observer. The distribution indicates that a fraction of the lensed sources at $z > 6$ have image separations $\lesssim 0.''1$.
Thus far, no multiple images have been detected with the HST in samples of $\sim 10$ $z>6$ quasars (e.g., \citealt{Kim_2009, McGreer_2013}). This is equivalent to stating that the fraction of $\mu > 2$ magnified quasars at $z > 6$ is $\lesssim 10\%$ in current samples, leading to an upper limit of $\beta \lesssim 3.1$ (see Fig. \ref{fig:theory}). It is worth noting that, while the diffraction limit for the HST is $\sim 0.''1$, its point-spread function (PSF) is very well characterized and stable. This would, in principle, allow to discern a point-source quasar from a multiply imaged one well below the diffraction limit, possibly at a level $\sim 0.''001$. For example, \cite{Libralato_2018} measured proper motions of stars by identifying the centroid of the PSF with a precision $\sim 0.''0003$. This remark could foster a re-examination of the HST images of $z>6$ quasars thus far detected. 

Additionally, as originally suggested by \cite{FAN_SCIENCE_PAPER}, the current selection criteria might have missed an important fraction of the quasar population at $z > 6$ because a lens galaxy in the same line of sight might have contaminated the drop-out band of the spectrum, leading to a misclassification of the source. 
It is remarkable to note that, if we de-magnified J0439+1634 by a factor $\sim 5$, reducing its total magnification to $\mu \sim 10$, its flux would have been comparable to the one from the foreground lens galaxy and, thus, the quasar would have been misclassified. This situation could be occurring systematically for $z>6$ quasars with $\mu \lesssim 10$. As the typical image separation for strongly lensed $z>6$ quasars is $\ll 1''$ and the typical angular dimension of a $z \sim 1$ lens galaxy is $\sim 1''$ (see Fig. \ref{fig:bias}), the probability of lensing is equivalent to the probability of having the lens galaxy and the quasar images within the same photometric aperture. The presence of contamination from the light of the lens galaxy is thus inevitable in strong lensing situations. It is worth noting that galaxies that are sufficiently massive ($M_{\star} \gtrsim \, 10^{10} \, \mathrm{M_{\odot}})$ to act as efficient gravitational lenses are most common at $z \lesssim 1$ \citep{Conselice_2016}. 

Next, we estimate the number of $z>6$ lensed quasars that could have been missed by current selection criteria. The following calculations are referred to a flux-limited survey with a flux limit of $z_{\mathrm{AB}} = 22.0$, in agreement with the \cite{Wang_2017} survey from which J0439+1634 was drawn.
Let ${\cal N}_{\rm obs}$ be the number of observed quasars, ${\cal N}_{\rm  obs, l}$ be the number of observed quasars that are lensed, ${\cal N}_{\rm \overline{obs}}$ be the number of quasars that are not observed, and be ${\cal N}_{\rm \overline{obs}, l}$ the number of quasars that are not observed and are lensed. We make the underlying assumption that the only cause for the lack of detection of this additional population of quasars is the fact that they are magnified, and hence the contamination of their drop-out spectra leads to their misclassification. This results in ${\cal N}_{\rm \overline{obs}} = {\cal N}_{\rm \overline{obs}, l}$. 
We define
\begin{equation}
\frac{{\cal N}_{\rm obs, l}}{{\cal N}_{\rm obs}} = P_{\rm obs} \, ,
\end{equation}
and
\begin{equation}
\frac{{\cal N}_{\rm obs, l} + {\cal N}_{\rm \overline{obs},l}}{{\cal N}_{\rm obs} + {\cal N}_{\overline{\rm obs}}} = P_0 \, .
\end{equation}
Based on our previous results, $P_{\rm obs} \approx 1/150$, while $P_0$ depends on the slope $\beta$. Solving for ${\cal N}_{\rm \overline{obs}}$:
\begin{equation}
{\cal N}_{\overline{\rm obs}} = {\cal N}_{\rm obs} \frac{P_0 - P_{\rm obs}}{1 - P_0} \, .
\end{equation}
We define the \textit{lensing boost factor} as
\begin{equation}
b_L = \frac{{\cal N}_{\rm \overline{obs}}}{{\cal N}_{\rm obs}} = \frac{P_0 - P_{\rm obs}}{1 - P_0} \, ,
\end{equation}
where $b_L$ is the fractional amount of undetected $z > 6$ quasars (normalized to the observed ones), which only depends on the overall probability of magnification $P_0$. 
Note that while a high magnification bias ${\cal B}$ favors the observation of more sources, a high lensing boost factor $b_L$ indicates that we are not observing a large fraction of the existing sources. If we had observed all of the objects that are predicted to be magnified ($P_{\rm obs} = P_0$), then $b_L=0$.
The lensing boost factor is the number by which the observed BH mass function of $z>6$ quasars, $\varphi_{\rm obs}(M_{\bullet}, z)$ needs to be corrected by a factor $(1 + b_L)$
\begin{equation}
\varphi_{\rm real}(M_{\bullet}, z) = (1 + b_L) \varphi_{\rm obs}(M_{\bullet}, z) \, .
\end{equation}

\begin{figure}
\includegraphics[angle=0,width=0.5\textwidth]{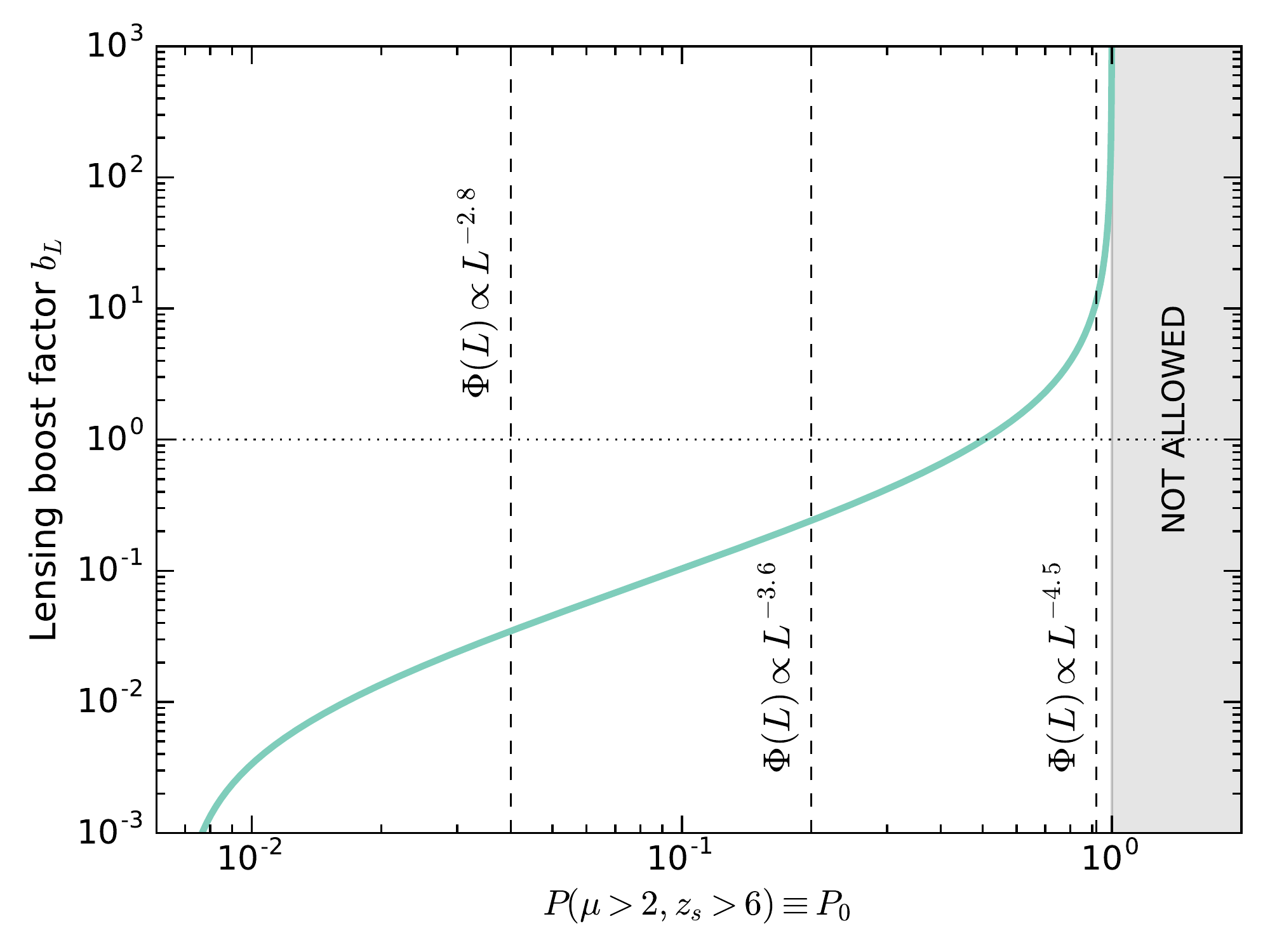}
\vspace{-0.4cm}
\caption{Lensing boost factor $b_L$ for the values of $P_0 = P(\mu > 2, z_s>6)$ considered in this Letter, $P_0 = 0.92$, $P_0 = 0.20$ and $P_0 = 0.04$. The horizontal line indicates the value of $P_0$ that makes the undetected population equal in number to the detected population. We assume a flux-limited survey with a flux limit of $z_{\mathrm{AB}} = 22.0$ \citep{Wang_2017}.}
\label{fig:booster}
\end{figure}

With the values of $P_0$ considered in this Letter, the lensing boost factor varies within the range $0.03 < b_L < 12$, as shown in Fig. \ref{fig:booster}.
Also note that any value of $P_0 \gtrsim 0.5$ would lead to a lensing boost factor $b_L > 1$.

With the value of $\beta \sim 5$ predicted by \cite{Kulkarni_2018}, \textit{we might be currently missing the vast majority of the $z > 6$ quasar population}. 
With ${\cal N}_{\mathrm{obs}} \approx 150$, the previous range reads
\begin{equation}
5 \lesssim {\cal N}_{\overline{\mathrm{obs}}} \lesssim 1800 \, .
\end{equation}
Considering the space number density of $z>6$ quasars from the COSMOS-Legacy survey $\Phi_{\mathrm{obs}} \sim 100 \, \mathrm{Gpc^{-3}}$ (with an X-ray luminosity $L_X > 10^{44.1} \, \mathrm{erg \, s^{-1}}$, \citealt{Cosmos_Legacy2016}) we might be missing a contribution
\begin{equation}
3 \, \mathrm{Gpc^{-3}} \lesssim \Phi_{\overline{\mathrm{obs}}} \lesssim 1200 \, \mathrm{Gpc^{-3}} \, .  
\end{equation}

\section{Discussion and Conclusions} 
\label{sec:disc_concl}
We have explored the theoretical consequences of the detection of the first lensed $z > 6$ quasar \citep{FAN_SCIENCE_PAPER}. Our results represent an advance relative to previous calculations (such as \citealt{Wyithe_Loeb_2002}) in several respects: (i) we employ an updated, observationally motivated cosmological distribution of lenses and sources, and (ii) we now have at our disposal an observational point to calibrate our calculations.

The probability of a source at $z=6.51$ being lensed with magnification $\mu > 50$ depends on the slope of the bright end of the quasar luminosity function, and we obtain $P(\mu > 50) = P_{\rm obs}$ for $\beta \approx 3.2$. A discrepancy between the frequentist approximation of the probability and the theoretical expectation can be explained by taking into account the magnification bias.

Depending on the slope $\beta$, current predictions for the lensing probability range within $0.04<P_0<0.92$, in striking contrast with the observed value of $P_{\rm obs} \approx 7\times 10^{-3}$. It is thus likely that the observed population of $z > 6$ quasars contains cases with $\mu \lesssim 10$.

Additionally, the current selection criteria might have missed a significant fraction of the $z > 6$ quasars, due to the contamination of their light by lens galaxies. Assuming that this undetected population coincides with the set of magnified objects at $z>6$, we predict that the unknown lensed quasars could account for up to about half of the currently detected population for $\beta \lesssim 3.6$. Remarkably, for the most recently suggested \citep{Kulkarni_2018} value of $\beta \sim 5$, the vast majority of the $z > 6$ lensed cases of quasars is still undetected.
It is important to note that our formalism predicts the fraction of $z>6$ that are undetected only because the magnifying effect of a lens galaxy introduces a contamination in their drop-out band. The presence of additional populations of quasars that are undetected for other reasons (e.g., obscuration, \citealt{Comastri_2015}, or inefficient accretion, \citealt{Pacucci_2017}) is not taken into account here. 

We have introduced the lensing boost factor $b_L$ to model in a compact way the effect of an undetected population of $z>6$ quasars on the corresponding BH mass functions: $\varphi_{\rm real}(M_{\bullet}, z) = (1 + b_L) \varphi_{\rm obs}(M_{\bullet}, z)$.
Future high-$z$ quasar surveys, employing improved selection criteria, will be instrumental to setting constraints on $b_L$ and, consequently, determining the corrected quasar luminosity functions.

Some discussion is due on how these predicted lensed $z>6$ quasars are classified. It is reasonable to argue that they are misclassified as low-$z$ galaxies. In fact, a contamination of the drop-out bands of $z>6$ quasars (outside the IGM transmission windows) would lead to a failure of the photo-$z$ determination, thus artificially decreasing the redshift and misclassifying the quasar as a galaxy. This is the case as long as the following two conditions are met: (i) the flux from the lens galaxy is comparable to or larger than that of the quasars, and (ii) no X-ray emission is detected from the quasar. The first condition is likely met as the lens is usually a massive elliptical galaxy at $z \lesssim 1.5$ \citep{Mason_2015} while the quasar is at $z>6$. The difference in luminosity distance would cause the flux from the quasar to be comparable to the flux of the galaxy, even if the former is a hundred times more intrinsically luminous than the latter. Regarding the second condition, it is worth noting that the presence of the lens galaxy in the same line of sight would artificially increase the total gas column density, thus decreasing the probability of an X-ray detection.

In this Letter we considered two possible solutions to the problem of the lack of detection of lensed $z>6$ quasars: (i) a fraction of the currently detected $z>6$ quasars is lensed, but their images are unresolved; (ii) the lensed $z>6$ quasars are misclassified into low-$z$ galaxies, due to contamination effects in their photometry.
These two solutions are complementary and not in conflict with each other. Understanding their feasibility requires advanced modeling of the lens population and, possibly, ad hoc modeling of some specific lensing configurations. In fact, the detectability of the lens galaxy ultimately depends on the ratio between its flux and the magnified flux of the quasar. More advanced models are required to provide a final answer to the question: how many lensed $z > 6$ quasars might we be missing?

An additional population of undetected $z > 6$ quasars could have implications to our understanding of the epoch of reionization. For instance, it could change our view of the contribution of quasars to the reionization of the Universe (e.g., \citealt{Madau_2017}).
Furthermore, there is a long-standing search for the origin of the infrared (e.g., \citealt{Kashlinsky_2007,Yue_2013}) and X-ray backgrounds (e.g., \citealt{Gilli_2007}) and their cross correlation \citep{Cappelluti_2013}. Our predicted population of undetected $z>6$ quasars, with their spectra extended from the infrared to the X-ray, could make an important contributions to these backgrounds.
Moreover, if a fraction of the currently detected population of $z>6$ quasars is found to be magnified, the value of their BH mass would be consequently decreased, easing the problem of early supermassive BH growth (e.g., \citealt{Volonteri_2005, Pacucci_2017, Pacucci_2018}).
For these reasons, employing new selection criteria to better understand the population of high-$z$ quasars will be of fundamental importance in the near future. Our results strongly highlight the importance of the discovery by \cite{FAN_SCIENCE_PAPER} and will guide future extended searches for lensed quasars.

\vspace{0.3cm}
We thank the anonymous referee for constructive comments on the manuscript.
F.P. acknowledges support from the NASA \textit{Chandra} award No. AR8-19021A and enlightening discussions with Xiaohui Fan, Massimo Meneghetti, and Andrea Ferrara.
This work was supported in part by the Black Hole Initiative at Harvard University, which is funded by a JTF grant.




\label{lastpage}
\end{document}